\documentclass[10pt,conference]{IEEEtran}

\usepackage{amsmath,amsfonts}
\usepackage{graphicx}
\usepackage{booktabs}
\usepackage{multirow}
\usepackage{hyperref}
\usepackage{xcolor}
\usepackage{academicons}  

\hypersetup{
    colorlinks=true,
    linkcolor=black,
    urlcolor=[HTML]{A6CE39},
    hidelinks
}

\usepackage{tikz}
\usetikzlibrary{
    arrows.meta,
    positioning,
    calc,
    backgrounds,
    decorations.pathreplacing
}

\usepackage{cuted}
\usepackage{capt-of}

\title{Magnitude Is All You Need? Rethinking Phase in Quantum Encoding of Complex SAR Data}

\author{
\IEEEauthorblockN{Sakthi Prabhu Gunasekar and Prasanna Kumar Rangarajan}
\IEEEauthorblockA{
Dept. of Computer Science \& Engineering\\
Amrita School of Computing\\
Amrita Vishwa Vidyapeetham, India\\
ORCID: \href{https://orcid.org/0009-0006-0153-5674}{0009-0006-0153-5674},
\href{https://orcid.org/0000-0001-6103-259X}{0000-0001-6103-259X}
}
}

\begin{document}

\maketitle
 
\begin{abstract}
Synthetic Aperture Radar (SAR) data is inherently complex-valued, while quantum machine learning (QML) models operate in complex Hilbert spaces. This similarity suggests that using both the magnitude and phase of the data in quantum encoding should help with tasks like automatic target recognition (ATR) in SAR images. In this study, we test this idea by comparing five different ways to encode the data for quantum models: using only magnitude, using both magnitude and phase together, using in-phase and quadrature components, using preprocessed phase, and using a pure quantum architecture. We test all these approaches using the same experimental setup on the MSTAR benchmark dataset.

Surprisingly, we find that magnitude-only encoding works better than phase-inclusive encodings in hybrid quantum-classical models. It achieves 99.57\% accuracy on the 3-class task and 71.19\% on the 8-class task, outperforming complex-valued alternatives under the same experimental framework. Adding phase information provides little or no improvement and can sometimes degrade performance. However, in purely quantum models with only 184--224 trainable parameters and no classical neural-network layers, phase information becomes much more important, improving accuracy by up to 21.65 percentage points.

These findings show that the usefulness of phase information depends not only on the data, but also on the architecture used to process it. Hybrid models can make up for missing phase information with their classical parts, but pure quantum models need phase information to tell classes apart. Our results give practical advice for encoding complex-valued data in quantum machine learning and show how important it is to design encoding and model architecture together, especially with current quantum technology.
\end{abstract}


\section{Introduction}
 
Synthetic Aperture Radar (SAR) serves as a valuable tool
for automatic target recognition (ATR) because it functions
reliably regardless of weather or lighting conditions~\cite{kechagias2020atr_survey}.
Unlike standard optical sensors, SAR captures data with both
magnitude and phase components, each offering unique insights
into the observed scene. The magnitude relates to the intensity
of the scattered signal and reveals details about the target's
shape and geometry, while the phase contains information about
the signal's travel path and the physical characteristics of
the reflecting materials~\cite{novak1997sar_atr}. Thus, these two
components provide complementary perspectives: magnitude
describes how prominent a reflection is, and phase offers
clues about the origin and nature of the signals through
interference patterns. Given that quantum states are inherently
described by both amplitude and phase, it might seem logical
that quantum models would benefit from including phase
information. However, our findings suggest that this advantage
is not always realized in practice.

In recent years, quantum machine learning (QML) has gained
attention as a powerful approach for handling complex
data-driven problems. QML exploits distinctive quantum
properties such as superposition, entanglement, and
interference to process information within high-dimensional
Hilbert spaces~\cite{biamonte2017qml,schuld2019qml}. Hybrid
quantum-classical architectures, in particular, have achieved
strong results on various image classification
benchmarks~\cite{cherrat2024qvit,singh2026medmnist,zhang2025hqvit}.
These models are especially notable in scenarios where quantum
circuits can match the accuracy of classical solutions while
using far fewer trainable parameters.

The inherent complex-valued nature of both SAR data and
quantum states suggests a natural alignment: quantum circuits
should be well-suited for processing SAR signals, and
incorporating both magnitude and phase into quantum encoding
should improve performance. However, this assumption has not
been systematically validated. Most QML studies focus on real-valued image data (e.g., MNIST, CIFAR), and the few quantum approaches applied to radar or remote sensing domains do not address the specific challenges of encoding complex-valued inputs~\cite{senokosov2023qml_image}.
 
In this work, we present the first systematic study of quantum encoding strategies for SAR ATR. We evaluate five encoding strategies under a unified experimental framework, namely magnitude-only, joint complex, I/Q decomposition, preprocessed phase, and pure quantum encoding. These strategies are tested on the MSTAR benchmark dataset~\cite{mstar_dataset} across three architectural variants, namely hybrid quantum-classical, dual-path quantum, and purely quantum models. We refer to our hybrid quantum-classical architecture with magnitude-only encoding as \textbf{MagQT} (Magnitude Quantum Transformer).
 
Our results reveal a surprising and previously unreported finding: \textbf{in hybrid quantum-classical architectures, magnitude-only encoding consistently outperforms all complex-valued strategies}, achieving 99.57\% accuracy on a 3-class task and 71.19\% on an 8-class task. Phase-inclusive methods provide small, negligible, or negative improvements in hybrid settings. However, \textbf{in purely quantum architectures, phase information becomes essential}, contributing up to +21.65\% improvement in accuracy.
 
This paradox reveals a fundamental interaction between encoding strategy and model design. Phase can be useful or redundant depending on the architectural context. Classical components in hybrid architectures compensate for missing phase information, whereas purely quantum models require phase to construct discriminative representations.

At a high level, our study answers a simple but fundamental question: does giving quantum models more information (phase) always help? Surprisingly, the answer depends not on the data, but on the architecture itself.
 
The main contributions of this work are:

\begin{enumerate}
\item \textbf{Phase paradox discovery:} We show that the utility of phase depends on the architecture. Phase is negligible in hybrid quantum-classical models but critical in purely quantum models. This reveals a previously unreported interaction between encoding strategy and model design.

\item \textbf{Systematic encoding study:} We perform the first comprehensive comparison of five encoding strategies for complex-valued data, demonstrating that magnitude-only encoding is optimal in hybrid architectures despite the complex-valued nature of SAR signals.

\item \textbf{Parameter-efficient hybrid quantum model (MagQT):} We introduce MagQT, a hybrid quantum transformer that achieves 99.57\% accuracy on 3-class MSTAR with only $\sim$58K parameters, matching classical CNN performance while requiring 5--188$\times$ fewer parameters, highlighting the potential of quantum feature extraction for parameter-efficient learning.

\item \textbf{Design guidelines for NISQ systems:} We establish practical principles for handling complex-valued data in quantum machine learning, including when phase information should be used and how classical preprocessing enables effective quantum representations.
\end{enumerate}

\section{Related Work}
 
\subsection{Quantum Machine Learning for Vision Tasks}
 
Quantum machine learning (QML) has emerged as a promising paradigm for leveraging quantum systems in data-driven tasks~\cite{biamonte2017qml}. Variational Quantum Circuits (VQCs)~\cite{cerezo2021variational, benedetti2019pqc} and hybrid quantum-classical models have been widely explored for image classification. Farhi and Neven~\cite{farhi2018classification} demonstrated classification with quantum neural networks on near-term processors, establishing the viability of parameterized quantum circuits for supervised learning. Abbas et al.~\cite{abbas2021power} analyzed the expressivity of quantum neural networks, showing that quantum models can achieve favorable capacity-to-parameter ratios compared to classical counterparts.
 
For vision tasks specifically, Cherrat et al.~\cite{cherrat2024qvit} proposed Quantum Vision Transformers (QViT) using quantum orthogonal layers for image classification on MedMNIST datasets. Zhang et al.~\cite{zhang2025hqvit} introduced HQViT, a hybrid quantum vision transformer that uses swap-test based attention to compute attention coefficients in quantum feature space. Singh et al.~\cite{singh2026medmnist} presented the first comprehensive benchmarking of QML models on MedMNIST using real IBM quantum hardware, achieving up to 85.4\% accuracy on PneumoniaMNIST with purely quantum circuits. Cong et al.~\cite{cong2019qcnn} proposed quantum convolutional neural networks that leverage quantum circuit structure for hierarchical feature extraction.
 
While these works demonstrate the potential of quantum models for image classification, they focus exclusively on real-valued image data and do not address the challenges of encoding complex-valued inputs into quantum circuits.
 
\subsection{SAR Automatic Target Recognition}
 
SAR ATR has been extensively studied using classical machine learning and deep learning~\cite{kechagias2020atr_survey}. The MSTAR benchmark dataset~\cite{mstar_dataset}, collected by DARPA/AFRL using an X-band sensor at 1-foot resolution, has served as the standard evaluation platform for over two decades. Classical CNNs achieve near-perfect accuracy ($>$99\%) on the standard 3-class MSTAR configuration~\cite{chen2016cnn_sar}, with recent transformer-based approaches further improving robustness under extended operating conditions. Shao et al.~\cite{shao2020cnn_comparison} compared classical CNN architectures on MSTAR, showing that most achieve $\sim$99\% accuracy on 10-class SOC. Fu et al.~\cite{fu2020resnet_sar} achieved 99.67\% using ResNet with center-softmax loss.
 
SAR data is inherently complex-valued, and several works have explored complex-valued neural networks to jointly process magnitude and phase. Zeng et al.~\cite{zeng2022cv_sar} proposed multi-stream complex-valued networks for SAR target recognition, demonstrating that complex-valued processing can improve feature extraction. Trabelsi et al.~\cite{trabelsi2018deep_complex} introduced deep complex networks that extend convolutional operations to the complex domain. However, the practical contribution of phase information to classification performance remains inconsistent in classical pipelines, with magnitude-only approaches often matching or exceeding complex-valued methods.
 
To our knowledge, no prior work systematically studies how magnitude, phase, I/Q, and pure quantum encodings affect SAR ATR performance across hybrid and purely quantum architectures.
 
\subsection{Encoding of Complex-Valued Data in QML}
 
How data is encoded plays a crucial role in determining how well quantum machine learning (QML) models perform~\cite{schuld2021encoding}. There are several ways to encode data, such as angle, amplitude, and basis encoding. Each method strikes its own balance between how efficiently it uses qubits and how much information it can represent~\cite{havlicek2019supervised}. For example, P\'{e}rez-Salinas et al.~\cite{perez2020data_reuploading} demonstrated that by reintroducing input features at different layers within a quantum circuit, a process known as data reuploading, even circuits with just one qubit are capable of approximating any function. This approach is a key part of the models we use in our work.

Although quantum systems are, by nature, able to represent complex numbers through the amplitudes and phases of their states, most real-world QML applications still depend on encodings that use only real numbers. Up to now, there has not been a thorough investigation into how the choice of encoding method interacts with different model structures, especially when it comes to the role of phase information in hybrid versus fully quantum models.
 
\subsection{Gap and Motivation}

Despite the natural compatibility between complex-valued SAR data and quantum representations, existing work has not systematically examined how different quantum encodings of SAR magnitude, phase, and I/Q components affect ATR performance. In particular, the role of phase information remains unclear across hybrid quantum-classical and purely quantum architectures. This work addresses this gap through a controlled experimental study on the MSTAR benchmark, comparing five encoding strategies under a unified framework.

\section{Quantum Encoding Strategies for Complex-Valued SAR Data}
 
The performance of quantum machine learning models is strongly influenced by how classical data is encoded into quantum states. In the context of SAR, each pixel is inherently complex-valued, represented as $x = a + jb$, where $a$ and $b$ denote the in-phase (I) and quadrature (Q) components. Alternatively, this can be expressed in polar form as magnitude $|x|$ and phase $\phi = \arg(x)$.
 
In this work, we systematically evaluate five encoding strategies that differ in how magnitude and phase information are incorporated into quantum circuits. All strategies are implemented within a unified framework to ensure a fair comparison.

\paragraph{Data Reuploading}
Several of our encoding strategies employ data reuploading~\cite{perez2020data_reuploading}, a technique in which input features are re-encoded into the quantum circuit at multiple layers rather than only once. This is essential when the number of input features exceeds the number of available qubits. For example, encoding 64 pooled phase features on 8 qubits requires 8 reuploading layers, with each layer encoding a different subset of 8 features. P\'{e}rez-Salinas et al.\ showed that single-qubit circuits with sufficient data reuploading layers can approximate arbitrary functions, making this technique critical for expressibility in qubit-limited NISQ settings.

\subsection{Magnitude-Only Encoding (S1)}
 
In this approach, only the magnitude component $|x|$ is used, while phase information is discarded. The magnitude values are normalized and encoded into quantum states using parameterized rotation gates:
\begin{equation}
|\psi\rangle = \prod_{i} R_y(\alpha \cdot |x_i|)\, |0\rangle
\end{equation}
where $\alpha$ is a scaling factor and $i$ indexes the input features (e.g., image patches). This strategy is simple, stable, and avoids the potential noise introduced by phase. It serves as the baseline for all comparisons. Intuitively, this acts as a noise-resistant representation, focusing only on stable structural information.

\subsection{Joint Complex Encoding (S2)}
 
In joint complex encoding, both magnitude and phase are encoded into the same qubit using two rotation gates:
\begin{equation}
|\psi\rangle = \prod_{i} R_z(\phi_i)\, R_y(|x_i|)\, |0\rangle
\end{equation}
Here, $R_y$ encodes the magnitude and $R_z$ encodes the phase. This approach directly maps the polar representation of SAR data into quantum rotations, leveraging both amplitude and phase of the quantum state. This tightly couples magnitude and phase on the same qubit, making the representation highly expressive but also sensitive to perturbations.

\subsection{I/Q-Based Encoding (S3)}
 
Instead of converting to magnitude and phase, the raw real and imaginary components are directly encoded:
\begin{equation}
|\psi\rangle = \prod_{i} R_y(a_i)\, R_z(b_i)\, |0\rangle
\end{equation}
where $a_i = |x_i|\cos(\phi_i)$ and $b_i = |x_i|\sin(\phi_i)$ represent the in-phase and quadrature components, respectively. This quantum-native encoding maps the complex SAR signal directly onto the Bloch sphere, with $R_y$ controlling the polar angle and $R_z$ controlling the azimuthal angle. This can be seen as a more ``quantum-native'' representation, directly mapping complex numbers onto the Bloch sphere geometry.

\subsection{Preprocessed Phase Encoding (S4)}
 
In this strategy, magnitude and phase are processed through separate pathways. Magnitude is encoded using standard quantum encoding (e.g., $R_y$ rotations), while phase is first preprocessed via spatial average pooling and then encoded into a dedicated quantum circuit:
\begin{equation}
|\psi\rangle = U_{\text{mag}}(|x|) \otimes U_{\text{phase}}(\phi)
\end{equation}
where $U_{\text{mag}}$ and $U_{\text{phase}}$ denote separate quantum transformations for magnitude and phase. The outputs are combined at a later stage via late fusion. By separating magnitude and phase, this approach attempts to balance stability (magnitude) and expressiveness (phase).

\subsection{Pure Quantum Encoding (S5)}
 
In the pure quantum setting, both magnitude and phase are encoded and processed entirely within quantum circuits without any classical feature extraction or transformation layers:
\begin{equation}
|\psi\rangle = U_{\theta}(|x|, \phi)\, |0\rangle
\end{equation}
where $U_{\theta}$ represents a variational quantum circuit with trainable parameters. Unlike hybrid approaches, the model relies solely on quantum transformations for feature learning and classification. In this case, the model has no classical fallback, forcing the quantum circuit to fully exploit all available information, including phase.

\subsection{Discussion}
 
These five strategies represent different design choices in handling complex-valued data within quantum models. Magnitude-only encoding simplifies the input representation, while joint and I/Q encodings attempt to preserve full complex information within a single quantum pathway. Preprocessed phase encoding separates the modalities, and pure quantum encoding removes classical components entirely. This systematic categorization enables us to isolate the impact of phase information and analyze how different encoding strategies interact with hybrid and purely quantum architectures.


\begin{figure}[t]
\centering
\resizebox{\columnwidth}{!}{%
\begin{tikzpicture}[
    font=\small,
    qgate/.style={draw, fill=blue!15, minimum width=7mm, minimum height=5.5mm, align=center, font=\scriptsize},
    tgate/.style={draw, fill=orange!20, minimum width=7mm, minimum height=5.5mm, align=center, font=\scriptsize},
    pgate/.style={draw, fill=green!18, minimum width=7mm, minimum height=5.5mm, align=center, font=\scriptsize},
    rgate/.style={draw, fill=red!12, minimum width=7mm, minimum height=5.5mm, align=center, font=\scriptsize},
    cnot/.style={draw, circle, fill=black, inner sep=1.2pt},
    oplus/.style={draw, circle, minimum size=4.5mm, font=\scriptsize},
    wire/.style={line width=0.5pt}
]

\node[draw=none, font=\bfseries\small] at (5.5, 1.0) {(a) S1: Magnitude-Only --- $R_y(|x|)$};

\foreach \y/\q in {0/0, -1.0/1, -2.0/2, -3.0/3, -4.0/4, -5.0/5} {
    \node[anchor=east, font=\scriptsize] at (-0.3, \y) {$q_{\q}$};
    \draw[wire] (0, \y) -- (10.5, \y);
    \node[anchor=west, font=\scriptsize] at (10.6, \y) {$\langle Z \rangle$};
}

\foreach \y in {0, -1.0, -2.0, -3.0, -4.0, -5.0} {
    \node[qgate] at (1.2, \y) {$R_y$};
}

\foreach \ya/\yb in {0/-1.0, -1.0/-2.0, -2.0/-3.0, -3.0/-4.0, -4.0/-5.0} {
    \node[cnot] at (2.8, \ya) {};
    \node[oplus] at (2.8, \yb) {$+$};
    \draw[wire] (2.8, \ya) -- (2.8, \yb);
}

\foreach \y in {0, -1.0, -2.0, -3.0, -4.0, -5.0} {
    \node[tgate] at (4.2, \y) {$R_y$};
}

\foreach \y in {0, -1.0, -2.0, -3.0, -4.0, -5.0} {
    \node[qgate] at (5.8, \y) {$R_y$};
}

\foreach \ya/\yb in {0/-1.0, -1.0/-2.0, -2.0/-3.0, -3.0/-4.0, -4.0/-5.0} {
    \node[cnot] at (7.2, \ya) {};
    \node[oplus] at (7.2, \yb) {$+$};
    \draw[wire] (7.2, \ya) -- (7.2, \yb);
}

\foreach \y in {0, -1.0, -2.0, -3.0, -4.0, -5.0} {
    \node[tgate] at (8.6, \y) {$R_y$};
}

\draw[decorate, decoration={brace, amplitude=5pt}] (0.5, 1.5) -- (4.8, 1.5)
    node[midway, above=6pt, font=\small\bfseries] {Layer 1 (reupload)};
\draw[decorate, decoration={brace, amplitude=5pt}] (5.2, 1.5) -- (9.2, 1.5)
    node[midway, above=6pt, font=\small\bfseries] {Layer 2 (reupload)};

\begin{scope}[yshift=-7.8cm]
\node[draw=none, font=\bfseries\small] at (5.5, 1.0) {(b) S3: I/Q Encoding --- $R_y(I)\, R_z(Q)$};

\foreach \y/\q in {0/0, -1.0/1, -2.0/2, -3.0/3, -4.0/4, -5.0/5} {
    \node[anchor=east, font=\scriptsize] at (-0.3, \y) {$q_{\q}$};
    \draw[wire] (0, \y) -- (10.5, \y);
    \node[anchor=west, font=\scriptsize] at (10.6, \y) {$\langle Z \rangle$};
}

\foreach \y in {0, -1.0, -2.0, -3.0, -4.0, -5.0} {
    \node[qgate] at (0.8, \y) {$R_y$};
}

\foreach \y in {0, -1.0, -2.0, -3.0, -4.0, -5.0} {
    \node[pgate] at (1.9, \y) {$R_z$};
}

\foreach \ya/\yb in {0/-1.0, -1.0/-2.0, -2.0/-3.0, -3.0/-4.0, -4.0/-5.0} {
    \node[cnot] at (3.2, \ya) {};
    \node[oplus] at (3.2, \yb) {$+$};
    \draw[wire] (3.2, \ya) -- (3.2, \yb);
}

\foreach \y in {0, -1.0, -2.0, -3.0, -4.0, -5.0} {
    \node[tgate] at (4.4, \y) {$R_y$};
}

\foreach \y in {0, -1.0, -2.0, -3.0, -4.0, -5.0} {
    \node[qgate] at (5.8, \y) {$R_y$};
}
\foreach \y in {0, -1.0, -2.0, -3.0, -4.0, -5.0} {
    \node[pgate] at (6.9, \y) {$R_z$};
}
\foreach \ya/\yb in {0/-1.0, -1.0/-2.0, -2.0/-3.0, -3.0/-4.0, -4.0/-5.0} {
    \node[cnot] at (8.0, \ya) {};
    \node[oplus] at (8.0, \yb) {$+$};
    \draw[wire] (8.0, \ya) -- (8.0, \yb);
}
\foreach \y in {0, -1.0, -2.0, -3.0, -4.0, -5.0} {
    \node[tgate] at (9.2, \y) {$R_y$};
}

\draw[decorate, decoration={brace, amplitude=5pt}] (0.2, 1.5) -- (5.0, 1.5)
    node[midway, above=6pt, font=\small\bfseries] {Layer 1};
\draw[decorate, decoration={brace, amplitude=5pt}] (5.2, 1.5) -- (9.8, 1.5)
    node[midway, above=6pt, font=\small\bfseries] {Layer 2};

\end{scope}

\begin{scope}[yshift=-16.0cm]
\node[draw=none, font=\bfseries\small] at (5.5, 1.2) {(c) S5: Phase VQC --- Full Data Reuploading (8q, showing 4)};

\foreach \y/\q in {0/0, -1.0/1, -1.8/{\vdots}, -2.6/6, -3.4/7} {
    \node[anchor=east, font=\scriptsize] at (-0.3, \y) {$q_{\q}$};
}
\foreach \y in {0, -1.0, -2.6, -3.4} {
    \draw[wire] (0, \y) -- (10.5, \y);
}
\draw[wire, dotted] (0, -1.8) -- (10.5, -1.8);

\foreach \y in {0, -1.0, -2.6, -3.4} {
    \node[anchor=west, font=\scriptsize] at (10.6, \y) {$\langle Z \rangle$};
}

\node[draw=none, font=\scriptsize, above] at (1.0, 0.35) {sub-1};
\foreach \y in {0, -1.0, -2.6, -3.4} {
    \node[qgate] at (1.0, \y) {$R_y$};
}

\node[cnot] at (2.0, 0) {};
\node[oplus] at (2.0, -1.0) {$+$};
\draw[wire] (2.0, 0) -- (2.0, -1.0);
\node[cnot] at (2.0, -2.6) {};
\node[oplus] at (2.0, -3.4) {$+$};
\draw[wire] (2.0, -2.6) -- (2.0, -3.4);

\node[draw=none, font=\scriptsize, above] at (3.2, 0.35) {sub-2};
\foreach \y in {0, -1.0, -2.6, -3.4} {
    \node[qgate] at (3.2, \y) {$R_y$};
}

\node[cnot] at (4.2, 0) {};
\node[oplus] at (4.2, -1.0) {$+$};
\draw[wire] (4.2, 0) -- (4.2, -1.0);
\node[cnot] at (4.2, -2.6) {};
\node[oplus] at (4.2, -3.4) {$+$};
\draw[wire] (4.2, -2.6) -- (4.2, -3.4);

\node[font=\Large] at (5.3, -1.7) {$\cdots$};

\node[draw=none, font=\scriptsize, above] at (6.5, 0.35) {sub-8};
\foreach \y in {0, -1.0, -2.6, -3.4} {
    \node[qgate] at (6.5, \y) {$R_y$};
}

\node[cnot] at (7.5, 0) {};
\node[oplus] at (7.5, -1.0) {$+$};
\draw[wire] (7.5, 0) -- (7.5, -1.0);
\node[cnot] at (7.5, -2.6) {};
\node[oplus] at (7.5, -3.4) {$+$};
\draw[wire] (7.5, -2.6) -- (7.5, -3.4);

\node[draw=none, font=\scriptsize, above] at (8.5, 0.35) {Train};
\foreach \y in {0, -1.0, -2.6, -3.4} {
    \node[tgate] at (8.5, \y) {$R_y$};
}
\foreach \y in {0, -1.0, -2.6, -3.4} {
    \node[rgate] at (9.5, \y) {$R_z$};
}

\draw[decorate, decoration={brace, amplitude=5pt, mirror}] (0.3, -4.2) -- (10.0, -4.2)
    node[midway, below=7pt, font=\small\bfseries] {One layer ($\times L$) --- all 64 features via 8 sub-layers};

\end{scope}

\begin{scope}[yshift=-21.5cm]

\node[qgate, minimum width=10mm, minimum height=7mm] at (0.8, 0) {$R_y$};
\node[draw=none, anchor=west, font=\small] at (1.5, 0) {Data encoding (mag / I)};

\node[pgate, minimum width=10mm, minimum height=7mm] at (5.5, 0) {$R_z$};
\node[draw=none, anchor=west, font=\small] at (6.2, 0) {Phase / Q encoding};

\node[tgate, minimum width=10mm, minimum height=7mm] at (9.5, 0) {$R_y$};
\node[draw=none, anchor=west, font=\small] at (10.2, 0) {Trainable};

\end{scope}

\end{tikzpicture}%
}
\caption{Quantum circuit diagrams for key encoding strategies. (a)~S1: Magnitude-only with $R_y$ rotations on 6~qubits with data reuploading. (b)~S3: I/Q encoding with $R_y(I)$ and $R_z(Q)$ on the same qubit. (c)~S5: Phase VQC with full data reuploading across 8 sub-layers per variational layer.}
\label{fig:circuits}
\end{figure}
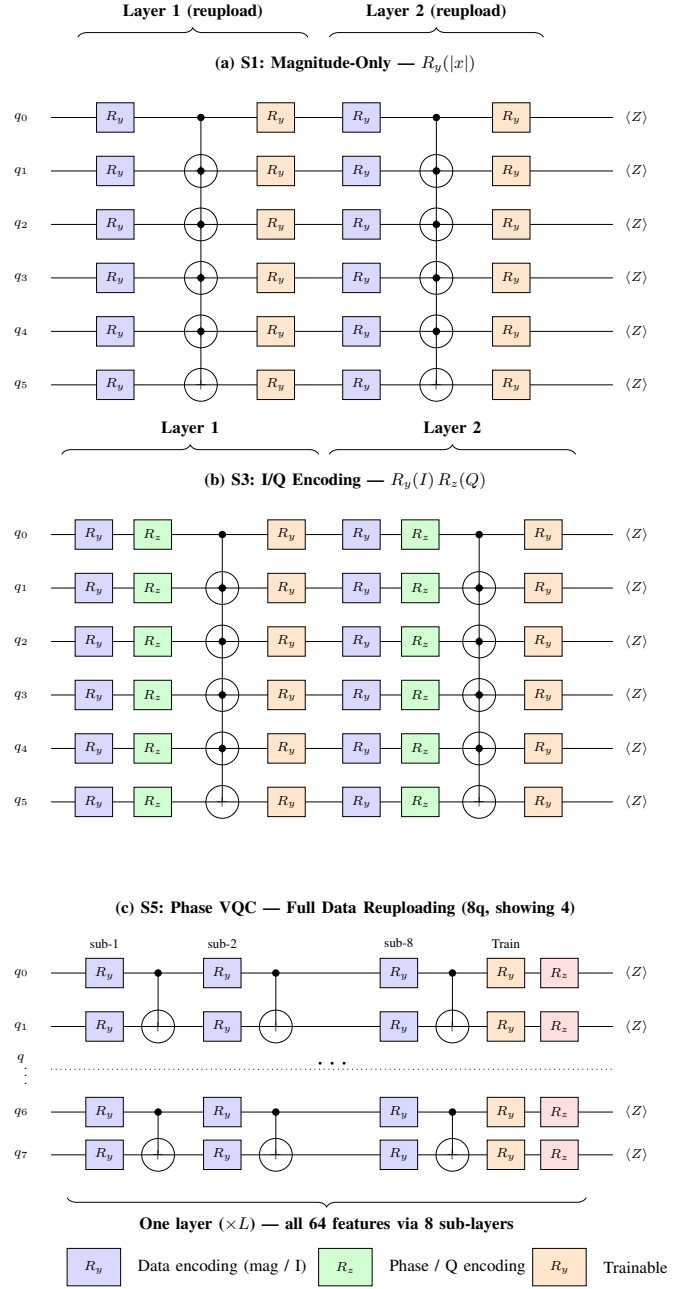

\section{Model Architecture}

Conceptually, the three architectures represent different levels of reliance on quantum vs classical computation. The hybrid model uses quantum circuits as feature extractors, the dual-path model explicitly separates signal components, and the pure quantum model relies entirely on quantum transformations for learning.
 
To enable a fair comparison of encoding strategies, we design a unified framework in which the downstream architecture remains consistent while only the input encoding varies. We consider three architectural variants: hybrid quantum-classical, dual-path quantum, and purely quantum models. All quantum circuits are implemented using PennyLane~\cite{bergholm2018pennylane} with the \texttt{lightning.qubit} backend and adjoint differentiation.

\subsection{Hybrid Quantum-Classical Architecture (MagQT)}
 
The hybrid model forms the primary backbone used in our experiments. We refer to this architecture with magnitude-only encoding (S1) as \textbf{MagQT} (Magnitude Quantum Transformer). The input SAR image (128$\times$128) is center-cropped to a 64$\times$64 region of interest (ROI) and divided into a 4$\times$4 grid of non-overlapping 16$\times$16 patches, yielding 16 patch tokens.

\paragraph{Quantum Patch Encoder}
For each patch, the top-$k$ pixels by magnitude intensity are selected ($k=6$), and their values are encoded into a 6-qubit parameterized quantum circuit. The circuit consists of $L=2$ variational layers, each comprising data encoding via $R_y$ rotations, a nearest-neighbor CNOT entangling chain, and trainable $R_y$ rotations:
\begin{equation}
U_l = \prod_{i} R_y(\theta_{l,i}) \cdot \text{CNOT}_{\text{chain}} \cdot \prod_{i} R_y(\alpha \cdot x_i)
\end{equation}
Pauli-$Z$ expectation values are measured on all 6 qubits, producing a 6-dimensional feature vector per patch, which is linearly projected to the embedding dimension $d=64$.
 
MagQT achieves 99.57\% accuracy on the 3-class MSTAR task with only $\sim$58K total parameters, including 12 quantum parameters and lightweight classical projection, transformer, and classifier layers. This matches classical CNN-level performance while using 5--188$\times$ fewer parameters. This parameter efficiency arises from the quantum encoder's ability to extract discriminative patch features in a 64-dimensional Hilbert space and return compact 6-dimensional expectation-value features, reducing the representational burden on the downstream classical transformer.

\paragraph{Transformer Block}
Patch embeddings are prepended with a learnable \texttt{[CLS]} token and augmented with positional embeddings. A single transformer block ($N_{\text{blocks}}=1$) applies self-attention followed by a feed-forward network (FFN) with GELU activation and dropout ($p=0.1$). The \texttt{[CLS]} output serves as the global image representation.

The complete hybrid pipeline is:
\begin{equation}
\mathbf{y} = f_{\text{cls}} \left( f_{\text{transformer}} \left( f_{\text{quantum}}(\mathbf{x}) \right) \right)
\end{equation}
 
\subsection{Dual-Path Quantum Architecture}
 
To explicitly evaluate the contribution of phase information, we introduce a dual-path architecture in which magnitude and phase are processed by separate quantum circuits.
 
\paragraph{Magnitude Path}
Identical to MagQT: 6-qubit patch encoder $\rightarrow$ transformer $\rightarrow$ \texttt{[CLS]} output (64-dim).

\paragraph{Phase Path}
The phase ROI (64$\times$64) is reduced via adaptive average pooling to 8$\times$8 = 64 features, normalized to $[0, \pi]$. These features are encoded into an 8-qubit variational quantum circuit using angle encoding with data reuploading across $L=8$ layers. Each layer encodes 8 features via $R_y$ rotations, applies a circular CNOT entangling ring, and applies trainable $R_y$ and $R_z$ rotations. Pauli-$Z$ measurements yield an 8-dimensional output, projected to 32 dimensions via a linear layer.
 
\paragraph{Late Fusion}
The magnitude \texttt{[CLS]} token (64-dim) and phase features (32-dim) are concatenated to form a 96-dimensional vector, which is passed through an MLP classifier:
\begin{equation}
\mathbf{z} = [\mathbf{z}_{\text{mag}} \| \mathbf{z}_{\text{phase}}], \quad \mathbf{y} = f_{\text{cls}}(\mathbf{z})
\end{equation}

\subsection{Pure Quantum Architecture}
 
In the purely quantum setting, all classical neural network components are removed. The model consists of four cascaded variational quantum circuits with zero classical trainable parameters.
 
\paragraph{Patch VQC} Each 16$\times$16 patch is average-pooled to 2$\times$2 = 4 features, normalized to $[0, \pi]$. A 4-qubit VQC with $L=4$ layers and full data reuploading encodes each patch, producing 4 Pauli-$Z$ measurements per patch ($16 \times 4 = 64$ total).
 
\paragraph{Global Magnitude VQC} The 64 patch-level measurements are re-normalized and fed into an 8-qubit VQC with $L=4$ layers and 8 sub-layers of full data reuploading per layer, producing 8 global magnitude features.
 
\paragraph{Phase VQC} The phase ROI is average-pooled to 8$\times$8 = 64 features, processed by an 8-qubit VQC with the same structure as the global magnitude VQC, producing 8 phase features.
 
\paragraph{Fusion VQC} The 16 concatenated features (8 magnitude + 8 phase) are fed into an $N_{\text{class}}$-qubit VQC ($N_{\text{class}} = 3$ or $8$) with $L=4$ layers. Pauli-$Z$ measurements on each qubit serve directly as class scores; the predicted class is obtained via $\arg\max$.

\subsection{Complexity Analysis}

Table~\ref{tab:complexity} summarizes the quantum resources and parameter counts across all three architectures. The hybrid model uses the fewest quantum parameters but relies heavily on classical components. The pure quantum model eliminates all classical parameters at the cost of reduced classification capacity.

\begin{table}[t]
\centering
\caption{Quantum resource and parameter comparison across architectures.}
\label{tab:complexity}
\resizebox{\columnwidth}{!}{
\begin{tabular}{lccccc}
\toprule
Component & Qubits & Layers & Q. Params & C. Params & Total \\
\midrule
\multicolumn{6}{l}{\textit{Hybrid Architecture --- MagQT (S1)}} \\
\quad Patch encoder & 6 & 2 & 12 & 3,676 & 3,688 \\
\quad Transformer & -- & 1 & -- & 50,112 & 50,112 \\
\quad Classifier & -- & -- & -- & 4,355 & 4,355 \\
\quad \textbf{Total} & \textbf{6} & & \textbf{12} & \textbf{$\sim$58K} & \textbf{$\sim$58K} \\
\midrule
\multicolumn{6}{l}{\textit{Dual-Path Architecture (S4)}} \\
\quad Patch encoder & 6 & 2 & 12 & 390 & 402 \\
\quad Phase VQC & 8 & 8 & 128 & 288 & 416 \\
\quad Transformer & -- & 1 & -- & 33,408 & 33,408 \\
\quad Classifier & -- & -- & -- & 26,730 & 26,730 \\
\quad \textbf{Total} & \textbf{14} & & \textbf{140} & \textbf{60,816} & \textbf{$\sim$61K} \\
\midrule
\multicolumn{6}{l}{\textit{Pure Quantum Architecture (S5, 8-class)}} \\
\quad Patch VQC & 4 & 4 & 32 & 0 & 32 \\
\quad Global mag VQC & 8 & 4 & 64 & 0 & 64 \\
\quad Phase VQC & 8 & 4 & 64 & 0 & 64 \\
\quad Fusion VQC & 8 & 4 & 64 & 0 & 64 \\
\quad \textbf{Total} & \textbf{28} & & \textbf{224} & \textbf{0} & \textbf{224} \\
\bottomrule
\end{tabular}
}
\end{table}

\subsection{Design Rationale}
This unified framework allows us to attribute differences
in model performance mainly to the data encoding strategies,
rather than to changes in model architecture. The hybrid
model takes advantage of classical components to handle
broader reasoning tasks, while the pure quantum model is
used to explore how well quantum circuits can represent
information on their own. By including a dual-path design, we
are able to specifically examine the role that phase information
plays. Having this range of model types makes it possible
to systematically investigate how various encoding methods
interact with different learning approaches, which is central to
our experimental analysis.


\begin{figure*}[t]
\centering
\resizebox{\textwidth}{!}{%
\begin{tikzpicture}[
    font=\small,
    >=Latex,
    arrow/.style={->, line width=1.0pt},
    darrow/.style={->, dashed, line width=1.0pt},
    cmod/.style={
        draw, fill=gray!8, rounded corners=3pt, line width=0.9pt,
        align=center, minimum height=13mm, minimum width=24mm, inner sep=3pt
    },
    qmod/.style={
        draw, fill=blue!8, double, double distance=1pt, rounded corners=3pt,
        line width=0.9pt, align=center, minimum height=13mm,
        minimum width=24mm, inner sep=3pt
    },
    encmod/.style={
        draw, fill=yellow!8, dashed, very thick, rounded corners=3pt,
        align=center, minimum height=14mm, minimum width=32mm, inner sep=4pt
    },
    dimlab/.style={
        font=\scriptsize\color{blue!70!black}, midway, above, inner sep=2pt
    },
    dimlabb/.style={
        font=\scriptsize\color{blue!70!black}, midway, below, inner sep=2pt
    }
]

\node[draw=none, font=\bfseries\normalsize] at (10.5, 5.8) {(a) Hybrid Quantum--Classical Architecture (S1, S2, S3)};

\node[cmod] (a_in) at (0.8, 4.0) {
\textbf{SAR Image}\\
\small 128$\times$128
};

\node[cmod] (a_crop) at (4.0, 4.0) {
\textbf{Preprocess}\\
\small ROI 64$\times$64\\
\small 4$\times$4 patches
};

\node[encmod] (a_enc) at (8.5, 4.0) {
\textbf{Encoding (swap)}\\
\small S1: $R_y(|x|)$\\
\small S2: $R_y(|x|)\,R_z(\phi)$\\
\small S3: $R_y(I)\,R_z(Q)$
};

\node[qmod] (a_qpatch) at (13.0, 4.0) {
\textbf{Patch VQC}\\
\small 6 qubits, $L$=2\\
\small 12 Q-params
};

\node[cmod] (a_proj) at (16.5, 4.0) {
\textbf{Project}\\
\small 6$\to$64
};

\node[cmod] (a_trans) at (19.5, 4.0) {
\textbf{Transformer}\\
\small 1 block, $d$=64\\
\small CLS + pos emb
};

\node[cmod] (a_cls) at (22.5, 4.0) {
\textbf{MLP Head}\\
\small 64$\to N_c$
};

\draw[arrow] (a_in) -- (a_crop) node[dimlab] {128$^2$};
\draw[arrow] (a_crop) -- (a_enc) node[dimlab] {16 patches};
\draw[arrow] (a_enc) -- (a_qpatch) node[dimlab] {top-6 px};
\draw[arrow] (a_qpatch) -- (a_proj) node[dimlab] {16$\times$6};
\draw[arrow] (a_proj) -- (a_trans) node[dimlab] {17$\times$64};
\draw[arrow] (a_trans) -- (a_cls) node[dimlab] {CLS: 64};

\node[draw=none, font=\small, align=center] at (10.5, 2.6) {
Total: $\sim$58K params (12 quantum + $\sim$58K classical)
};

\node[draw=none, font=\bfseries\normalsize] at (10.5, 1.9) {(b) Dual-Path Phase-Aware Architecture (S4)};

\node[cmod] (b_in) at (0.8, -0.5) {
\textbf{SAR Image}\\
\small 128$\times$128
};

\node[cmod] (b_split) at (4.0, -0.5) {
\textbf{ROI + Split}\\
\small mag / phase
};

\node[encmod] (b_magenc) at (8.0, 0.8) {
\textbf{Mag Encoding}\\
\small $R_y(|x|)$, top-6
};

\node[qmod] (b_qpatch) at (12.0, 0.8) {
\textbf{Patch VQC}\\
\small 6q, $L$=2
};

\node[cmod] (b_trans) at (15.5, 0.8) {
\textbf{Transformer}\\
\small CLS $\to$ 64
};

\node[cmod] (b_pool) at (8.0, -2.0) {
\textbf{Avg Pool}\\
\small 64$^2$$\to$8$\times$8
};

\node[qmod] (b_phvqc) at (12.0, -2.0) {
\textbf{Phase VQC}\\
\small 8q, $L$=8\\
\small data reupload
};

\node[cmod] (b_phproj) at (15.5, -2.0) {
\textbf{Project}\\
\small 8$\to$32
};

\node[cmod] (b_fuse) at (19.5, -0.5) {
\textbf{Concat + MLP}\\
\small [64$\|$32] $\to N_c$
};

\draw[arrow] (b_in) -- (b_split);
\draw[arrow] (b_split) |- (b_magenc);
\draw[arrow] (b_magenc) -- (b_qpatch) node[dimlab] {top-6};
\draw[arrow] (b_qpatch) -- (b_trans) node[dimlab] {16$\times$64};
\draw[arrow] (b_trans) -| (b_fuse) node[dimlab, pos=0.3] {64};

\draw[darrow] (b_split) |- (b_pool);
\draw[darrow] (b_pool) -- (b_phvqc) node[dimlabb] {64 feats};
\draw[darrow] (b_phvqc) -- (b_phproj) node[dimlabb] {8};
\draw[darrow] (b_phproj) -| (b_fuse) node[dimlabb, pos=0.3] {32};

\node[draw=none, font=\small, align=center] at (10.5, -3.3) {
Total: $\sim$61K params (140 quantum + $\sim$61K classical)
};

\node[draw=none, font=\bfseries\normalsize] at (10.5, -4.0) {(c) Pure Quantum Architecture (S5) --- Zero Classical NN Params};

\node[cmod] (c_in) at (0.8, -6.0) {
\textbf{SAR Image}\\
\small 128$\times$128
};

\node[cmod] (c_crop) at (4.0, -6.0) {
\textbf{ROI + Pool}\\
\small mag: 16$\times$4\\
\small phase: 64
};

\node[qmod] (c_pvqc) at (8.0, -5.2) {
\textbf{Patch VQC}\\
\small 4q, $L$=4\\
\small $\times$16 patches\\
\small reupload
};

\node[qmod] (c_gvqc) at (12.0, -5.2) {
\textbf{Global VQC}\\
\small 8q, $L$=4\\
\small 8 sub-layers\\
\small reupload
};

\node[qmod] (c_phvqc) at (8.0, -7.2) {
\textbf{Phase VQC}\\
\small 8q, $L$=4\\
\small 8 sub-layers\\
\small reupload
};

\node[qmod] (c_fuse) at (16.0, -6.0) {
\textbf{Fusion VQC}\\
\small $N_c$ qubits, $L$=4\\
\small reupload
};

\node[cmod] (c_meas) at (19.5, -6.0) {
\textbf{Measure}\\
\small $\langle Z_i \rangle$$\to$argmax
};

\draw[arrow] (c_in) -- (c_crop);
\draw[arrow] (c_crop) |- (c_pvqc) node[dimlab, pos=0.3] {16$\times$4};
\draw[arrow] (c_pvqc) -- (c_gvqc) node[dimlab] {64};
\draw[arrow] (c_gvqc) -- (c_fuse) node[dimlab] {8};

\draw[darrow] (c_crop) |- (c_phvqc) node[dimlabb, pos=0.3] {64};
\draw[darrow] (c_phvqc) -| (c_fuse) node[dimlabb, pos=0.3] {8};

\draw[arrow] (c_fuse) -- (c_meas) node[dimlab] {$N_c$};

\node[draw=none, font=\small, align=center] at (10.5, -8.3) {
Total: 224 params (ALL quantum, zero classical NN parameters)
};

\node[qmod, minimum width=14mm, minimum height=8mm] (legq) at (3.0,-9.5) {};
\node[draw=none, anchor=west, font=\small] at (4.0,-9.5) {Quantum module};

\node[cmod, minimum width=14mm, minimum height=8mm] (legc) at (9.5,-9.5) {};
\node[draw=none, anchor=west, font=\small] at (10.5,-9.5) {Classical module};

\node[encmod, minimum width=14mm, minimum height=8mm] (lege) at (16.0,-9.5) {};
\node[draw=none, anchor=west, font=\small] at (17.0,-9.5) {Encoding (variable)};

\node[draw=none, font=\small] at (22.0,-9.5) {
\textcolor{blue!70!black}{Blue}: dimensions
};

\end{tikzpicture}%
}
\caption{Unified architecture overview. (a)~Hybrid quantum-classical architecture (MagQT) supporting S1--S3 via swappable encoding. (b)~Dual-path architecture (S4) with separate magnitude and phase quantum pathways. (c)~Pure quantum architecture (S5) with zero classical parameters. Dimensions shown in blue.}
\label{fig:arch_overview}
\end{figure*}
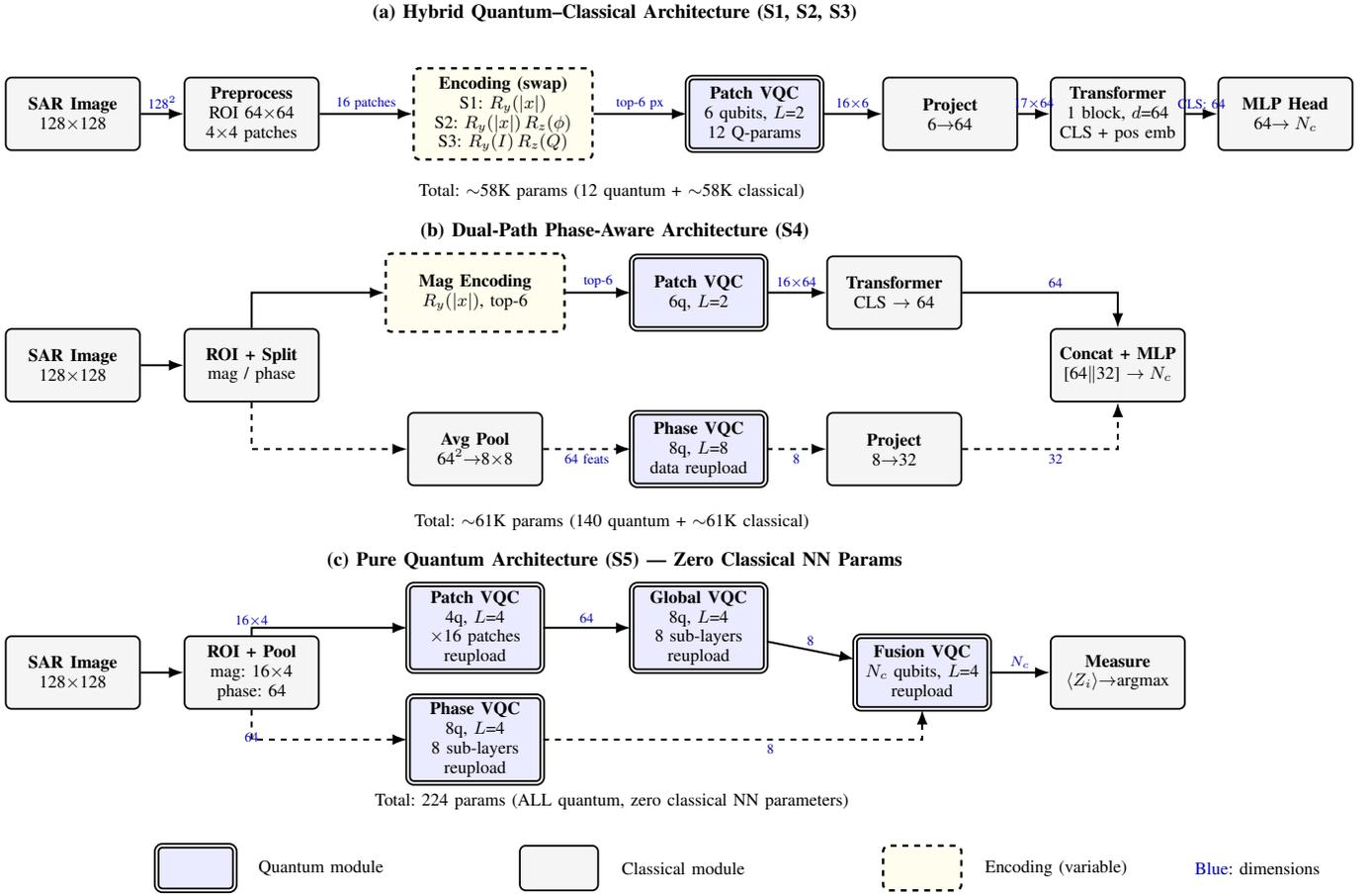
\section{Experiments}
 
\subsection{Dataset}
 
We evaluate all encoding strategies on the Moving and Stationary Target Acquisition and Recognition (MSTAR) benchmark dataset~\cite{mstar_dataset}, collected by DARPA and the Air Force Research Laboratory (AFRL) using an X-band SAR sensor in spotlight mode at 1-foot resolution. The dataset provides complex-valued SAR image chips (magnitude + phase) of military ground vehicles at multiple azimuth angles.

\begin{table}[t]
\centering
\caption{Dataset configuration for both classification tasks.}
\label{tab:dataset}
\begin{tabular}{lccc}
\toprule
Setting & Classes & Train & Test \\
\midrule
3-Class & 3 & 1,296 & 1,409 \\
8-Class (balanced) & 8 & 1,864 & 3,086 \\
\bottomrule
\end{tabular}
\end{table}

\begin{figure*}[t]
\centering
\includegraphics[width=\textwidth]{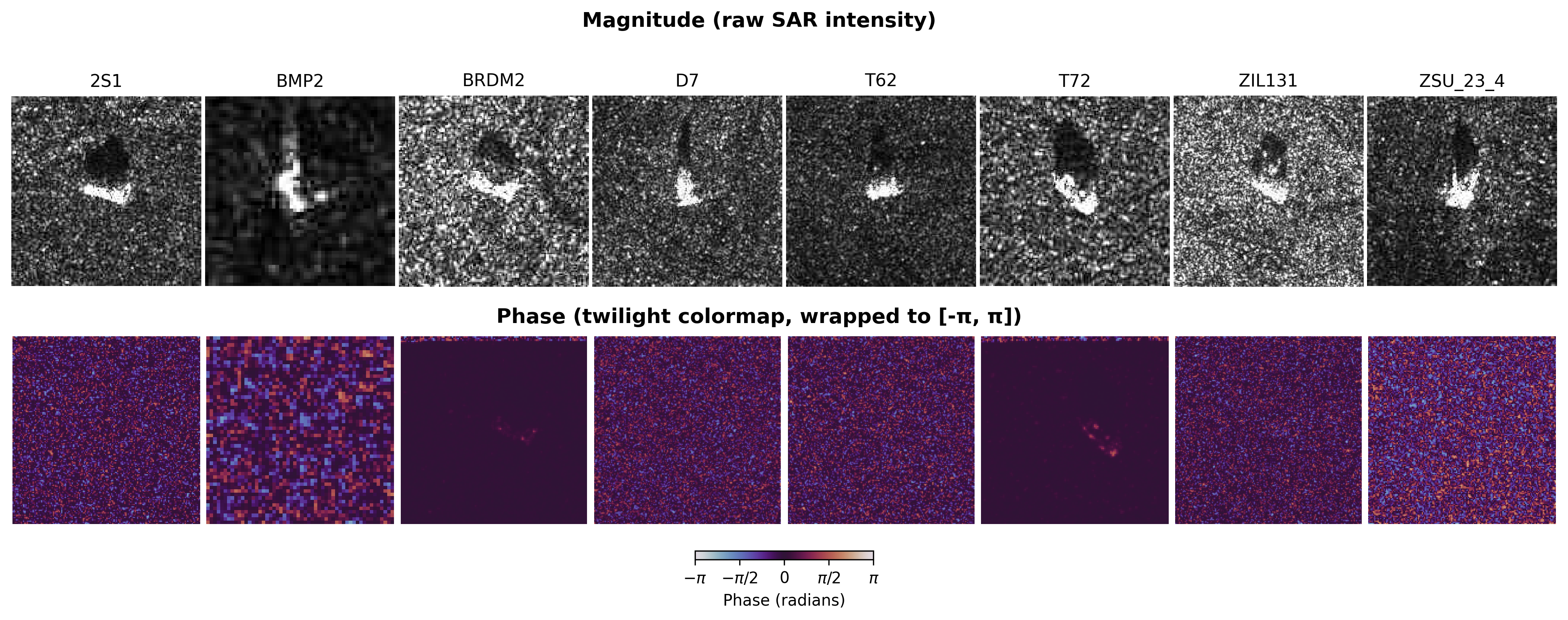}
\caption{MSTAR SAR samples for all 8 classes (ROI 64$\times$64 center crop). Top row: magnitude images showing distinct target structures. Bottom row: corresponding phase images, which appear noisy and visually indistinguishable across classes, motivating our investigation into whether quantum models can extract discriminative features from phase.}
\label{fig:mstar_samples}
\end{figure*}
 
We consider two classification settings:
 
\paragraph{3-Class Task} Three vehicle types: BMP2 (infantry fighting vehicle), ZIL131 (truck), and ZSU\_23\_4 (anti-aircraft gun). Training images are captured at 17$^\circ$ depression angle; testing at 15$^\circ$, following the standard operating condition (SOC) protocol.
 
\paragraph{8-Class Task} Eight vehicle types: 2S1, BMP2, BRDM2, D7, T62, T72, ZIL131, and ZSU\_23\_4. We exclude BTR60 and BTR70, which are visually near-identical armored personnel carriers known to cause persistent confusion in SAR ATR literature~\cite{kechagias2020atr_survey}. Training data is balanced at 233 samples per class to eliminate class imbalance effects.

\subsection{Preprocessing}
 
All input images (128$\times$128) are center-cropped to a 64$\times$64 region of interest (ROI). MSTAR targets are well-centered by design, and we empirically verified that center cropping significantly outperforms dynamic cropping approaches. Magnitude values are normalized to $[0, 1]$ and scaled by $\pi$ before quantum encoding. Phase values are normalized to $[0, \pi]$ for angle-based encodings. For I/Q encoding (S3), in-phase and quadrature components are computed as $I = |x| \cos(\phi)$ and $Q = |x| \sin(\phi)$.
 
\subsection{Implementation Details}
 
All quantum circuits are implemented in PennyLane~\cite{bergholm2018pennylane} using the \texttt{lightning.qubit} simulator with adjoint differentiation for gradient computation. Models are trained using PyTorch with the Adam optimizer ($\text{lr} = 10^{-4}$), cosine annealing learning rate schedule ($\eta_{\min} = 10^{-6}$), and cross-entropy loss with class weighting for the 8-class task. Training runs for 40 epochs with batch size 16. Each image is processed individually through the quantum circuits (per-sample forward pass), as batched quantum circuit execution is not natively supported.
 
\subsection{Main Results}
 
Table~\ref{tab:main_results} summarizes the performance of all encoding strategies. The phase contribution ($\Delta$) is measured via controlled ablation (Section~\ref{sec:ablation}).
 
\begin{table}[t]
\centering
\caption{Performance comparison of encoding strategies on MSTAR dataset. Hybrid QT denotes a hybrid quantum-classical transformer.}
\label{tab:main_results}
\resizebox{\columnwidth}{!}{
\begin{tabular}{lcccccc}
\toprule
Strategy & Type & Params & 3C Acc(\%) & 8C Acc(\%) & $\Delta_{3C}(\%)$ & $\Delta_{8C}$(\%) \\
\midrule
\textbf{S1: MagQT} & Hybrid QT & 58K & \textbf{99.57} & \textbf{71.19} & +0.00 & +0.03 \\
S4: Dual-path & Hybrid QT & 61K & 91.00 & 64.45 & +0.07 & -0.03 \\
S3: I/Q Encoding & Hybrid QT & 59K & 89.00 & 53.37 & +0.21 & -0.13 \\
\textbf{S5: Pure Quantum} & Pure Q & 184/224 & 70.62 & 37.01 & \textbf{+11.50} & \textbf{+21.65} \\
S2: Joint Complex & Hybrid QT & 59K & 67.49 & 59.10 & +1.30 & -0.60 \\
\bottomrule
\end{tabular}
}
\end{table}

From a high-level perspective, the results reveal a counterintuitive trend: adding more information (phase) does not necessarily improve performance. Instead, simpler representations often lead to better optimization and generalization in hybrid models.
 
\paragraph{Hybrid Architectures}
MagQT (S1) achieves the best performance among all hybrid encoding strategies, with 99.57\% accuracy on the 3-class task and 71.19\% on the 8-class task. All phase-inclusive hybrid strategies perform worse than the magnitude-only setting. The dual-path phase-aware model (S4) achieves 91.00\% on the 3-class task and 64.45\% on the 8-class task. I/Q encoding (S3) achieves 89.00\% and 53.37\%, respectively, while joint complex encoding (S2) achieves 67.49\% on the 3-class task and 59.10\% on the 8-class task. These results show that adding phase information does not improve hybrid quantum-classical SAR recognition relative to magnitude-only encoding under our experimental setting.
\paragraph{Pure Quantum Architecture}
In contrast, the pure quantum model (S5) exhibits a strong dependence on phase information, despite achieving lower overall accuracy than the hybrid models. Phase contributes +11.50\% on the 3-class task and +21.65\% on the 8-class task. With only 184--224 trainable parameters, all of which are quantum parameters, S5 achieves 70.62\% accuracy on the 3-class task and 37.01\% on the 8-class task. These results indicate that quantum circuits alone can learn discriminative SAR representations above random chance, although their limited capacity constrains performance on more complex multi-class settings.

\subsection{Comparison with Classical Methods}
 
Table~\ref{tab:classical_comparison} compares MagQT with established classical methods on MSTAR. While classical CNNs remain substantially stronger on the 10-class SOC task in absolute accuracy, MagQT uses far fewer trainable parameters and establishes a compact quantum-classical baseline for SAR ATR. On the 3-class task, MagQT reaches classical-level accuracy with a substantially smaller model. Notably, the pure quantum model (S5) achieves 70.62\% on the 3-class task with only 184 trainable parameters and no classical neural network layers, highlighting the parameter efficiency of the quantum-only setting.
 
\begin{table}[t]
\centering
\caption{Comparison with classical methods on MSTAR (SOC, 17$^\circ$$\to$15$^\circ$).}
\label{tab:classical_comparison}
\resizebox{\columnwidth}{!}{
\begin{tabular}{llccl}
\toprule
Method & Type & Params & Acc (\%) & Notes \\
\midrule
\multicolumn{5}{l}{\textit{Classical Methods (10-class)}} \\
ResNet~\cite{fu2020resnet_sar} & CNN & $\sim$11M & 99.67 & Center+softmax loss \\
MS-CNN~\cite{shao2020cnn_comparison} & Transfer & $\sim$5M & 98.83 & Transfer learning \\
A-ConvNets~\cite{chen2016cnn_sar} & CNN & $\sim$303K & 99.13 & Lightweight SAR CNN \\
Atrous-Inception & CNN & $\sim$3M & 97.97 & Inception variant \\
\midrule
\multicolumn{5}{l}{\textit{Our Quantum Methods (10-class)}} \\
\textbf{MagQT (S1)} & Hybrid QT & \textbf{58K} & \textbf{69.57} & Quantum transformer \\
S4: Dual-path & Hybrid QT & 61K & 61.92 & Phase VQC added \\
S1: Balanced & Hybrid QT & 59K & 57.35 & Class-balanced \\
\midrule
\multicolumn{5}{l}{\textit{Our Quantum Methods (3-class)}} \\
\textbf{MagQT (S1)} & Hybrid QT & \textbf{58K} & \textbf{99.57} & \textbf{Matches classical} \\
S5: Pure Quantum & Pure Q & \textbf{184} & 70.62 & Zero classical params \\
\bottomrule
\end{tabular}
}
\end{table}

\subsection{Ablation Study: Phase Contribution}
\label{sec:ablation}
 
\paragraph{Methodology}
To quantify the contribution of phase information, we perform a controlled test-time ablation: all phase-related inputs are replaced with zero vectors while model weights trained with phase remain unchanged. The validity of this approach depends on the architectural design:
 
\begin{itemize}
    \item \textbf{Late-fusion architectures (S1, S4):} The magnitude pathway operates independently of the phase pathway. Zeroing phase inputs does not affect magnitude processing, ensuring that any accuracy difference is attributable solely to the absence of phase information.
    
   \item \textbf{Pure quantum architecture (S5):} Magnitude and phase are processed through entirely separate quantum pathways. The magnitude path (Patch VQC $\rightarrow$ Global VQC) produces 8 magnitude features, while the Phase VQC independently produces 8 phase features. These 16 features are concatenated and processed by the Fusion VQC. During ablation, only the phase input is zeroed, producing the input vector $[\mathbf{z}_{\text{mag}}, \mathbf{0}]$ to the Fusion VQC. The magnitude features pass through unmodified. The observed accuracy drop ($-11.50\%$ on 3-class, $-21.65\%$ on 8-class) therefore reflects genuine reliance on phase information learned by the Fusion VQC, as the magnitude pathway remains fully intact.
 
    \item \textbf{Early-fusion architecture (S2):} S2 achieves 67.49\% accuracy on 3-class and 59.1\% on 8-class with phase contributions of +1.3\% and -0.6\% respectively, consistent with other hybrid architectures where phase provides negligible benefit.
\end{itemize}
 
\begin{table}[t]
\centering
\caption{Phase contribution across architectural paradigms.}
\label{tab:phase_ablation}
\begin{tabular}{lcc}
\toprule
Architecture & $\Delta_{\text{phase}}$ (3C) & $\Delta_{\text{phase}}$ (8C) \\
\midrule
Hybrid QT (MagQT) & $\approx 0\%$ & $\approx 0\%$ \\
Pure Q & $+11.50\%$ & $+21.65\%$ \\
\bottomrule
\end{tabular}
\end{table}

\subsection{Pure Quantum Analysis}
 
Intuitively, pure quantum models operate under a stricter constraint because they do not include classical layers that can suppress difficult or noisy features. As a result, they are forced to utilize phase information, even when it is unstable.

Although pure quantum models underperform hybrid architectures in absolute accuracy, they reveal important properties of quantum representations for SAR data. With only 184 parameters for the 3-class task and 224 parameters for the 8-class task, and with zero classical neural network layers, S5 achieves 70.62\% and 37.01\% accuracy, respectively. Both results are above random chance, which is 33.3\% for the 3-class task and 12.5\% for the 8-class task. This demonstrates that variational quantum circuits possess sufficient inductive bias for SAR target classification.
 
More significantly, the strong phase contribution in S5 (+11.50\% and +21.65\%) shows that SAR phase carries genuine discriminative signal that quantum circuits can exploit. Without phase, the 8-class pure quantum model drops to 15.36\%, which is only slightly above the random-chance level of 12.5\%. This demonstrates that phase is not merely helpful but essential in the pure quantum setting. These results suggest that as quantum hardware scales, architectures with greater quantum expressibility may be able to leverage phase while maintaining competitive classification accuracy.
 
\subsection{10-Class Results}
 
We additionally report results on the full 10-class task (Table~\ref{tab:10class}), where performance is bounded by the structural similarity of BTR60/BTR70 vehicles.
 
\begin{table}[t]
\centering
\caption{Results on full 10-class MSTAR task.}
\label{tab:10class}
\begin{tabular}{llccc}
\toprule
Strategy & Type & Params & Test Acc & $\Delta_{\text{phase}}$ \\
\midrule
\textbf{S1: MagQT} & Hybrid QT& 58K & \textbf{69.57\%} & +1.12\% \\
S4: Dual-path & Hybrid QT& 61K & 61.92\% & +0.52\% \\
S1: Balanced & Hybrid QT& 59K & 57.35\% & -0.06\% \\
\bottomrule
\end{tabular}
\end{table}

\section{Analysis: Why Does Magnitude Dominate?}

The key question is: why does a richer representation (magnitude + phase) perform worse in hybrid models? The answer lies in the interaction between information content and optimization stability.
 
The experimental results reveal a counterintuitive behavior: magnitude-only encoding consistently outperforms complex-valued strategies in hybrid architectures, while phase information becomes critical in purely quantum models. In this section, we analyze the underlying reasons for this phenomenon.
 
\subsection{Impact of Phase on Quantum Encoding Stability}
 
Quantum encoding works by turning classical features into
rotation angles within quantum circuits. Magnitude values are
usually smooth and stable, but phase values in SAR data can
be very sensitive to noise, wrapping, and minor changes in the
signal.
When phase is encoded using rotation gates like $R_z(\phi)$,
small changes in phase can cause the quantum state to oscillate
quickly. This results in unstable feature representations and can
make the learning process harder. On the other hand, encoding only magnitude leads to smoother, more stable changes in the quantum state, making optimization easier. In simple
terms, phase acts like a very sensitive signal: tiny differences
can greatly affect the quantum state and disrupt learning.
In contrast, changes in magnitude create smoother shifts,
allowing optimization algorithms to work more effectively.
 
\subsection{Gradient Analysis}
 
The training logs support the encoding stability argument
quantitatively. For the dual-path S4 run, phase VQC gradient
norms decline from nearly $0.5$ to under $0.02$ within the
first $10$ epochs, while magnitude encoder gradients remain
active throughout the training run, consistent within the
$0.5$--$2.0$ range. The optimizer routes information along
the dominant magnitude pathway and leaves the phase branch
largely dormant.

The pure quantum models (S5) suggest otherwise. Every circuit
gradient, phase VQC included, remains non-zero, holding
$0.1$--$2.5$ activity over all $40$ epochs. Since there is no
classical fallback, phase must be used. Combining these runs
shows hybrid architectures depending on magnitude as primary
and treating phase as secondary during training.
 
\subsection{Role of Classical Components in Hybrid Models}

In hybrid quantum-classical models, the quantum component
primarily acts as a feature extractor, with the classical
transformer and classifiers taking on the heavier
representation learning. These classical layers are capable
of recovering discriminative structure from the magnitude
alone, which compensates for information lost when phase
is excluded. This behavior matches classical SAR ATR results,
where magnitude-only CNNs routinely achieve $99\%$ accuracy
on MSTAR~\cite{chen2016cnn_sar}.
 
\subsection{Phase as a Critical Resource in Pure Quantum Models}
 
In purely quantum architectures, all representation learning
directly occurs on the quantum circuit without classical
layers to distribute the burden. Phase information feeds directly into quantum state interference, which
expands the expressive ability of variational circuits. The
ablation confirms this: stripping phase from the pure quantum
model drops three-class accuracy from $70.62\%$ to $59.12\%$
($-11.50\%$), and 8-class accuracy from $37.01\%$ to $15.36\%$
($-21.65\%$).

\subsection{The Information--Noise Trade-off}
 
The results suggest a fundamental trade-off: while phase carries additional information, it also introduces noise and instability. In hybrid architectures, the cost of noise outweighs the informational benefit. In purely quantum models, the informational benefit outweighs the noise cost because the model has no alternative source of discriminative features.

\begin{figure}[t]
\centering
\includegraphics[width=\columnwidth]{phase_paradox_ieee.png}
\caption{Phase contribution across architectures and encoding strategies. 
In hybrid quantum-classical models, phase contributions remain small, 
negligible, or negative and do not outperform magnitude-only encoding. 
In purely quantum models, phase becomes essential, contributing +11.50\% 
(3-class) and +21.65\% (8-class). This reveals that phase utility is 
architecture-dependent, not data-dependent.}
\label{fig:phase_paradox}
\end{figure}

\subsection{Implications for Quantum Machine Learning Design}
 
These findings lead to three practical design principles for quantum encoding of complex-valued data:
\begin{enumerate}
    \item \textbf{Encoding-architecture co-design}: The optimal encoding strategy cannot be determined independently of the model architecture. Phase inclusion should be considered only when classical components are absent or limited.
    \item \textbf{Classical preprocessing for noisy modalities}: When phase must be encoded, classical preprocessing such as spatial averaging should precede quantum encoding to reduce noise.
    \item \textbf{NISQ-pragmatic encoding}: In the current NISQ era, simpler encodings that avoid noisy inputs lead to more stable optimization and better performance.
\end{enumerate}
 
Overall, our analysis reveals that the utility of phase information is an emergent property of the interaction between encoding and architecture, rather than an inherent advantage of complex-valued representation.

\section{Conclusion}
 
In this work, we presented a systematic study of quantum encoding strategies for complex-valued SAR data under a unified experimental framework. By evaluating five encoding strategies across hybrid, dual-path, and purely quantum architectures, we arrive at a central finding: \textbf{magnitude-only encoding consistently outperforms all complex-valued strategies in hybrid quantum-classical architectures}, achieving 99.57\% accuracy on a 3-class task and 71.19\% on an 8-class task with fewer than 60K parameters.
 
More importantly, we uncover a fundamental paradox. Phase information is useful, but current NISQ-scale hybrid architectures do not reliably exploit it. Pure quantum circuits can leverage phase and achieve up to +21.65\% accuracy improvement, but they lack sufficient representational capacity to solve complex classification tasks. Hybrid architectures possess the capacity for high accuracy, but they render phase largely redundant because classical components compensate for the missing information.
 
Although pure quantum models do not achieve the highest absolute accuracy, they help uncover important aspects of how phase information is represented. These insights are hidden in hybrid systems. The sizable accuracy gains from phase in pure quantum settings
($+11.50\%$ on 3-class, $+21.65\%$ on 8-class) indicate that
SAR phase carries genuine class-discriminative signal. The
implication for future architecture work is direct: as
hardware moves past current NISQ constraints, deeper and more
expressive quantum designs should be able to exploit phase
without giving up classification accuracy.

Our findings push back on the assumption that complex-valued
encoding is uniformly beneficial in quantum machine learning.
The practical takeaways are threefold: (1) for today's hybrid
architectures, magnitude-only encoding is the better default;
(2) when phase is encoded, classical preprocessing should
precede it to suppress noise; and (3) encoding strategy and
model architecture need to be selected jointly rather than as
independent choices.

\subsection{Limitations}
 
This study has some important limitations. All experiments
were performed using quantum simulators, not real quantum
hardware, so the results could change when tested on actual
devices that introduce noise. The MSTAR dataset is widely
used as a benchmark, but it comes from controlled imaging
conditions and may not capture the full range of real-world
SAR situations. Furthermore, our pure quantum models use
relatively shallow circuits with fewer than 250 parameters,
which restricts their ability to represent complex patterns. A further limitation is that several experiments are reported from single training runs due to the high computational cost of per-sample quantum circuit simulation; multi-seed evaluation remains an important direction for estimating statistical variance.
 
\subsection{Future Work}
 
Several directions remain open for follow-up work: (1)
running these methods on physical quantum devices, following
the protocol used by Singh et al.~\cite{singh2026medmnist},
paired with error mitigation and correction; (2) scaling pure
quantum models to greater depth and qubit count to test
whether phase utility grows with circuit complexity; (3)
designing encoding schemes that adapt to phase noise at
encoding time; and (4) extending the evaluation to larger and
more challenging SAR datasets beyond MSTAR.

\section*{Acknowledgment}

The authors used Claude (Anthropic) for language editing, paraphrasing, and structural refinement assistance on selected portions of this manuscript, including Sections~II-A, IV-D, and VI-A--VI-E. All technical contributions, including model design, experiments, training, result verification, and analysis, are the authors' own original work. The authors reviewed and edited all AI-assisted text and take full responsibility for the content of this paper.



\begin{thebibliography}{99}
 
\bibitem{kechagias2020atr_survey}
O. Kechagias-Stamatis and N. Aouf, ``Automatic target recognition on synthetic aperture radar imagery: A survey,'' \textit{IEEE Access}, vol. 9, pp. 40562--40596, 2021.
 
\bibitem{novak1997sar_atr}
L. M. Novak, G. J. Owirka, W. S. Brower, and A. L. Weaver, ``The automatic target-recognition system in SAIP,'' \textit{Lincoln Laboratory Journal}, vol. 10, no. 2, pp. 187--202, 1997.
 
\bibitem{biamonte2017qml}
J. Biamonte, P. Wittek, N. Pancotti, P. Rebentrost, N. Wiebe, and S. Lloyd, ``Quantum machine learning,'' \textit{Nature}, vol. 549, no. 7671, pp. 195--202, 2017.
 
\bibitem{schuld2019qml}
M. Schuld and N. Killoran, ``Quantum machine learning in feature Hilbert spaces,'' \textit{Physical Review Letters}, vol. 122, no. 4, p. 040504, 2019.
 
\bibitem{cherrat2024qvit}
E. A. Cherrat \textit{et al.}, ``Quantum vision transformers,'' \textit{Quantum}, vol. 8, p. 1265, 2024.
 
\bibitem{singh2026medmnist}
G. Singh, H. Jin, and K. M. Merz Jr., ``Benchmarking MedMNIST dataset on real quantum hardware,'' \textit{Scientific Reports}, vol. 16, Article no. 9017, 2026.
 
\bibitem{zhang2025hqvit}
H. Zhang, Q. Zhao, M. Zhou, and L. Feng, ``HQViT: Hybrid quantum vision transformer for image classification,'' \textit{arXiv preprint arXiv:2504.02730}, 2025.
 
\bibitem{senokosov2023qml_image}
A. Senokosov \textit{et al.}, ``Quantum machine learning for image classification,'' \textit{arXiv preprint arXiv:2304.09224}, 2024.
 
\bibitem{mstar_dataset}
T. D. Ross, S. W. Worrell, V. J. Velten, J. C. Mossing, and M. L. Bryant, ``Standard SAR ATR evaluation experiments using the MSTAR public release data set,'' in \textit{Proc. SPIE}, vol. 3370, pp. 566--573, 1998.
 
\bibitem{bergholm2018pennylane}
V. Bergholm \textit{et al.}, ``PennyLane: Automatic differentiation of hybrid quantum-classical computations,'' \textit{arXiv preprint arXiv:1811.04968}, 2018.
 
\bibitem{schuld2021encoding}
M. Schuld, R. Sweke, and J. J. Meyer, ``Effect of data encoding on the expressive power of variational quantum-machine-learning models,'' \textit{Physical Review A}, vol. 103, no. 3, p. 032430, 2021.
 
\bibitem{perez2020data_reuploading}
A. P\'{e}rez-Salinas, A. Cervera-Lierta, E. Gil-Fuster, and J. I. Latorre, ``Data re-uploading for a universal quantum classifier,'' \textit{Quantum}, vol. 4, p. 226, 2020.
 
\bibitem{abbas2021power}
A. Abbas \textit{et al.}, ``The power of quantum neural networks,'' \textit{Nature Computational Science}, vol. 1, no. 6, pp. 403--409, 2021.
 
\bibitem{havlicek2019supervised}
V. Havl\'{i}\v{c}ek \textit{et al.}, ``Supervised learning with quantum-enhanced feature spaces,'' \textit{Nature}, vol. 567, no. 7747, pp. 209--212, 2019.
 
\bibitem{chen2016cnn_sar}
S. Chen, H. Wang, F. Xu, and Y.-Q. Jin, ``Target classification using the deep convolutional networks for SAR images,'' \textit{IEEE Trans. Geosci. Remote Sens.}, vol. 54, no. 8, pp. 4806--4817, 2016.
 
\bibitem{cong2019qcnn}
I. Cong, S. Choi, and M. D. Lukin, ``Quantum convolutional neural networks,'' \textit{Nature Physics}, vol. 15, no. 12, pp. 1273--1278, 2019.
 
\bibitem{farhi2018classification}
E. Farhi and H. Neven, ``Classification with quantum neural networks on near term processors,'' \textit{arXiv preprint arXiv:1802.06002}, 2018.
 
\bibitem{benedetti2019pqc}
M. Benedetti, E. Lloyd, S. Sack, and M. Fiorentini, ``Parameterized quantum circuits as machine learning models,'' \textit{Quantum Science and Technology}, vol. 4, no. 4, p. 043001, 2019.
 
\bibitem{zeng2022cv_sar}
Z. Zeng, J. Sun, Z. Han, and W. Hong, ``SAR automatic target recognition method based on multi-stream complex-valued networks,'' \textit{IEEE Trans. Geosci. Remote Sens.}, vol. 60, p. 5228618, 2022.
 
\bibitem{trabelsi2018deep_complex}
C. Trabelsi \textit{et al.}, ``Deep complex networks,'' \textit{arXiv preprint arXiv:1705.09792}, 2018.
 
\bibitem{cerezo2021variational}
M. Cerezo \textit{et al.}, ``Variational quantum algorithms,'' \textit{Nature Reviews Physics}, vol. 3, no. 9, pp. 625--644, 2021.
 
\bibitem{shao2020cnn_comparison}
J. Shao, C. Qu, and J. Li, ``A performance analysis of convolutional neural network models in SAR target recognition,'' in \textit{Proc. IGARSS}, pp. 1142--1145, 2017.
 
\bibitem{fu2020resnet_sar}
K. Fu, X. Zhang, and G. Wang, ``SAR ATR based on ResNet with center-softmax loss,'' in \textit{Proc. IEEE Int. Geosci. Remote Sens. Symp. (IGARSS)}, 2020.
 
\end{thebibliography}
\end{document}